\documentclass[twocolumn,tighten]{aastex631} 
\usepackage{multirow}
\usepackage{amsmath}
\usepackage{enumitem}
\usepackage{hyperref}

\setlist[description]{font=\normalfont\itshape}

\begin{document}

\title{Aliasing from Galactic Plane Setting in Widefield Radio Interferometry}

\author[0000-0003-2064-6979]{N. Barry}
\affiliation{International Centre for Radio Astronomy Research, Curtin University, Bentley, WA 6102, Australia}
\affiliation{ARC Centre of Excellence for All Sky Astrophysics in 3 Dimensions (ASTRO 3D), Australia}

\author[0000-0002-9130-5920]{J.~L.~B. Line}
\affiliation{International Centre for Radio Astronomy Research, Curtin University, Bentley, WA 6102, Australia}
\affiliation{ARC Centre of Excellence for All Sky Astrophysics in 3 Dimensions (ASTRO 3D)}

\author[0000-0002-0494-192X]{C.~R. Lynch}
\affiliation{Department of Physics and Astronomy, University of North Carolina Asheville, Asheville, NC 28804, USA}

\author[0000-0002-5270-6908]{M. Kriele}
\affiliation{International Centre for Radio Astronomy Research, University of Western Australia, Crawley, WA 6009, Australia}
\affiliation{International Centre for Radio Astronomy Research, Curtin University, Bentley, WA 6102, Australia}

\author[0000-0001-8001-0791]{J. Cook}
\affiliation{International Centre for Radio Astronomy Research, Curtin University, Bentley, WA 6102, Australia}
\affiliation{ARC Centre of Excellence for All Sky Astrophysics in 3 Dimensions (ASTRO 3D)}

\shorttitle{Aliasing from Galactic Plane Setting}
\shortauthors{Barry et. al}

\begin{abstract}

Measurements with widefield radio interferometers often include the near-infinite gradient between the sky and the horizon. This causes aliasing inherent to the measurement itself, and is purely a consequence of the Fourier basis. For this reason, the horizon is often attenuated by the instrumental beam down to levels deemed inconsequential. However, this effect is enhanced via our own Galactic plane as it sets over the course of a night. We show all-sky simulations of the Galactic plane setting in a low-frequency radio interferometer in detail for the first time. We then apply these simulations to the Murchison Widefield Array to show that a beam attenuation of 0.1\% is not sufficient in some precision science cases. We determine that the noise statistics of the residual data image are drastically more Gaussian with aliasing removal, and explore consequences in simulation for cataloging of extragalactic sources and 21 cm Epoch of Reionization detection via the power spectrum. 

\end{abstract}


\section{Introduction}

Survey science and all-sky statistics benefit from widefield observations, increasing the sky coverage and reducing signal-to-noise. In radio interferometry, many instruments now see a significant portion of the sky to reap these advantages. Some of the widest field-of-view instruments include the Murchison Widefield Array (MWA\footnote{19.4\,deg at 200\,MHz; \citealt{tingay_murchison_2013}}), the Hydrogen Epoch of Reionization Array (HERA\footnote{9\,deg at 150\,MHz; \citealt{deboer_hydrogen_2017}}), the New Extension in Nançay Upgrading LOFAR (Nenufar\footnote{7.1\,deg at 85\,MHz; \citealt{zarka_low-frequency_2020}}), the Low Frequency Array (LOFAR\footnote{6.38\,deg at 200\,MHz; \citealt{van_haarlem_lofar:_2013}}), and the Hydrogen Intensity and Real-time Analysis Experiment (HIRAX\footnote{5--10\,deg at 400--800\,MHz; \citealt{crichton_hydrogen_2022}}), to name a few.

However, by increasing the amount of sky seen per observation, many assumptions that built the field of radio interferometry are no longer satisfactory. More recently, the sharpness of the opaque horizon has been seen to affect widefield observations. Interferometers natively measure in a basis akin to Fourier space, and thus can experience aliasing at infinite gradients. The horizon represents an infinite gradient between sky emission and opaque ground, and bright emission at the horizon can exacerbate the amount of aliasing produced. 

Foregrounds at the horizon have been seen in Fourier space, creating a ``pitchfork" effect in the Fourier transform along element--element separations known as delay space \citep{thyagarajan_confirmation_2015}. Calibration accuracy can be improved by filtering the measurements themselves in time as foregrounds move across the night for drift-scan instruments \citep{charles_use_2023}. Noninterferometric radio observations of the full sky, like in global-signal experiments, see the horizon to a significant degree and thus must account for horizon emission in their models \citep{bassett_lost_2021}. In both delay-space corrections and global-signal models, statistical methods were used to indirectly suppress or observe horizon effects.

We present the most detailed image-space observations and simulations of the setting of the Galactic plane to definitively show the effect of horizon-based aliasing along with its progenitor. Our own Galaxy is an extremely bright source of synchrotron emission, e.g., over 2000\,Jy at 159\,MHz at the Galactic Center \citep{kriele_imaging_2022}, and subtends an arc over the whole sky. We show that as it sets over the opaque horizon, even at extremely low levels of beam sensitivity, it creates aliasing that is inherent to the observations themselves and is present throughout the entirety of the image. This has consequences in calibration, image generation, and noise characterization for precision science cases in widefield radio interferometry. 

We explore the effects of aliasing of the Galactic plane for two separate science cases: extragalactic catalogs and the Epoch of Reionization (EoR) detection.
\begin{description}

\item[Extragalactic catalogs] Creation of extragalactic catalogs requires deep images to determine flux density and shape of compact and diffuse sources. Even with advanced deconvolution strategies and a range of local sidereal time measurements, the aliasing of the Galactic plane will not fully decorrelate in image space and could affect the completeness of catalogs in certain regions of sky. 

\item[EoR] The redshifted 21 cm line from hydrogen during the EoR is expected to be extremely faint in comparison to the extragalactic and Galactic foregrounds; however they naturally separate in Fourier space due to their differing spectral signatures. Thus, measurements of this time period are pursued in the power spectrum or similar. Noise-like additions to the power spectrum that do not decorrelate with time will preclude the faint EoR measurement. 
\end{description}

In Section~\ref{sec:sims}, we simulate the effects of the setting of the Galactic plane in image space at 180\,MHz for interferometric observations, and then narrow our focus to simulated MWA Phase I observations of one of the coldest patches in the Southern hemisphere. In Section~\ref{sec:data}, we show how our simulations can mitigate the setting of the Galactic plane in real data and improve the overall noise statistics of the image. We propagate these effects into (1) the cataloging metric of catalog completeness and (2) the power spectrum metric of the 21 cm detection of the EoR in Section~\ref{sec:consequences}. Section~\ref{sec:analysis_and_data} outlines our methodologies and Section~\ref{sec:conclusion} summarizes our results.

%%%%%%%%%%%%%%%%%%%%%%%%%%%%%%%%%
%%%%%%%%%%%%%%%%%%%%%%%%%%%%%%%%%
\section{Analysis and Data}
\label{sec:analysis_and_data}
To show the effects of the setting of the Galactic plane in radio interferometry, we employ precise simulations of Galactic maps, horizon-to-horizon imaging software, and example data from the MWA. We highlight the key procedures and aspects of these highly complex frameworks.

\subsection{Catalogs and MWA Data}

The MWA is a radio interferometer that is composed of simple cross-dipoles arranged in square stations of 16. For the Phase I of the MWA (2013--2016; \citealt{tingay_murchison_2013}), there were 128 of these stations in a pseudo-random configuration. The maximum distance between stations (baseline) was 3\,km with a nearly filled aperture in Fourier space to about 50\,wavelengths. 

Our example data from the MWA was observed on 2014 July 1 at frequencies 167--198\,MHz. The target field was EoR0, centered on RA 0\,h, decl.~--27\,deg. Data is integrated over 2\,s intervals for a total time of 2\,min per observation. Over the course of the night, the EoR0 field is observed in a drift-and-shift mode, where the pointing center is updated every 30\,min to keep the field in view, and thus a small range of local sidereal times (LSTs) is covered every night. 

We use three catalogs for simulation and calibration purposes of these MWA data. 
\begin{description}

\item[Extragalactic] We use the bespoke LoBES catalog, which was built specifically for the precise requirements of EoR science with the MWA \citep{lynch_mwa_2021}. It uses a combination of point-source, Gaussian, and shapelet components to model 80,824 sources in 3069\,deg$^2$. LoBES is 70\% complete at 10.5 mJy and 90\% complete at 32 mJy.
\item[Cas~A] Near the horizon, we also observe Casseiopia~A (Cas~A), an extremely bright and extended source with flux density of at $\sim$\,18,500\,Jy at 74\,MHz \citep{perley_accurate_2017}. We use the supernova remnant modeling framework built by \citet{cook_investigating_2022} to build a Gaussian component model using high-resolution data from the Very Large Array (VLA). 
% The Galactic plane is bright and extended, reaching temperatures of 8,400\,K at 159\,MHz. 
\item[Galactic Plane] We use the catalog from \citet{kriele_imaging_2022}, natively measured at 159\,MHz using the Engineering Development Array 2 (EDA2; \citealt{wayth_engineering_2021}) since it most accurately represents the data. Due to a high gradient in beam sensitivity near the horizon, we transform the spherical harmonics to a high-resolution (1$^\prime$.7) series of point sources. We only include components within Galactic longitudes of $\pm$\,15\,deg to simulate the plane only.

\end{description}

\subsection{Simulation Framework}
\label{subsec:framework}
In total, we need to simulate over 6.7\,million components for each of the 56 time steps and 384 frequency channels per observation with near-perfect representation at the horizon. We use a CUDA-enabled simulator, \texttt{WODEN}\footnote{\url{https://github.com/JLBLine/WODEN}} \citep{line_woden_2022}, to simulate visibilities directly from the measurement equation: 
\begin{multline}
    V(u,v,w) = \iint \frac{dl\,dm\,dn}{\sqrt{1-n}} \,A(l,m,n) I(l,m,n) \\ e^{-2 \pi i (ul+vm+w(n-1))},
    \label{eq:me}
\end{multline}
where $\{l,m,n\}$ are directional cosines, and their Fourier-dual is baseline separation in wavelengths, $\{u,v,w\}$. The flux density of the model components, $I(l,m,n)$, are taken from the catalog. The beam sensitivity at any given point on the sky, $A(l,m,n)$, is calculated for each component's location using \texttt{hyperbeam}\footnote{\url{https://github.com/MWATelescope/mwa_hyperbeam}}.

These simulated visibilities are then used directly to make images. There are no added effects (e.g.,~calibration errors or noise) unless otherwise stated. In Section~\ref{subsec:ps}, we briefly mention the results of expected calibration errors for the 21 cm EoR science case---for this, we use a simple linear least squares solver between the simulated visibilities and reference visibilities to generate per-tile gains. We note that this is only to generate basic understanding, as calibration algorithms are generally much more complex.

To generate images, we use Fast Holographic Deconvolution\footnote{\url{https://github.com/EoRImaging/FHD}} (FHD; \citealt{sullivan_fast_2012,barry_fhd/eppsilon_2019}). FHD's resource requirement scales effectively, allowing efficient generation of horizon-to-horizon images. FHD also includes beam estimation algorithms, which are critical to reducing analysis systematics at the horizon with the MWA.

The beam, used for both gridding visibilities and image contouring, is estimated using a Gaussian decomposition of the MWA sensitivity as described by \citet{barry_role_2022}. This beam is fully analytic in the image and Fourier plane. There are no discrete fast Fourier transform effects in our Fourier plane caused by transforming an instrumentally accurate beam, which is particularly helpful for low sensitivity regions such as the horizon.

We grid each visibility onto a Fourier plane using the Gaussian decomposition estimation as a kernel. We also separately grid the contribution of the beam estimation to generate weights in the Fourier plane. The fast Fourier transform of these discrete planes then generates the orthogonal projection of the image in uniform weighting. Mask contours are generated at the 1\% level of the beam. No image-based cuts or padding is performed in order to keep propagated weights for the power spectrum (Section~\ref{subsec:ps}).

\subsection{Data Framework}
\label{subsec:data_framework}

Our framework for analyzing data in this work, whilst extremely similar to the simulation framework, has some key differences.

We must remove instrumental effects through calibration for observed visibilities. First, we capture per-frequency structure from the measured auto-correlations, which are very stable in time for the MWA, for each tile. Coherent noise is mitigated via an overall scaling to cross-correlations.

We use the aforementioned simulations in Section~\ref{subsec:framework} in a linear least squares solver to calculate the per-tile phases, and then fit low-order polynomials to the results. In general, the simulations include over 38,000 sources from the LoBES catalog, a Cas A model, and $\pm$\,15\,deg of the Galactic plane.

This combination of auto-correlation and cross-correlation information reduces the dependence on knowledge of the sky structure. For a more detailed description of calibration, please see \citet{barry_fhd/eppsilon_2019,li_first_2019}.

Gridding and imaging is treated the exact same as simulations in Section~\ref{subsec:framework}.

%%%%%%%%%%%%%%%%%%%%%%%%%%%%%%%%%
%%%%%%%%%%%%%%%%%%%%%%%%%%%%%%%%%
\section{Interferometric Galactic Plane Simulations}
\label{sec:sims}

Aliasing in the image space is dependent not only on the location/brightness of the setting source, but also the properties of the measuring interferometer. 
Therefore, we will first simulate how this aliasing occurs using a simple instrument, one that is both unpolarized and unattenuated by a beam response. We will then simulate how the aliasing is affected by a more realistic instrumental response via the MWA polarized beams. 

\subsection{Unpolarized, unattenuated interferometer}

\begin{figure*}
\centering
	\includegraphics[width =\textwidth]{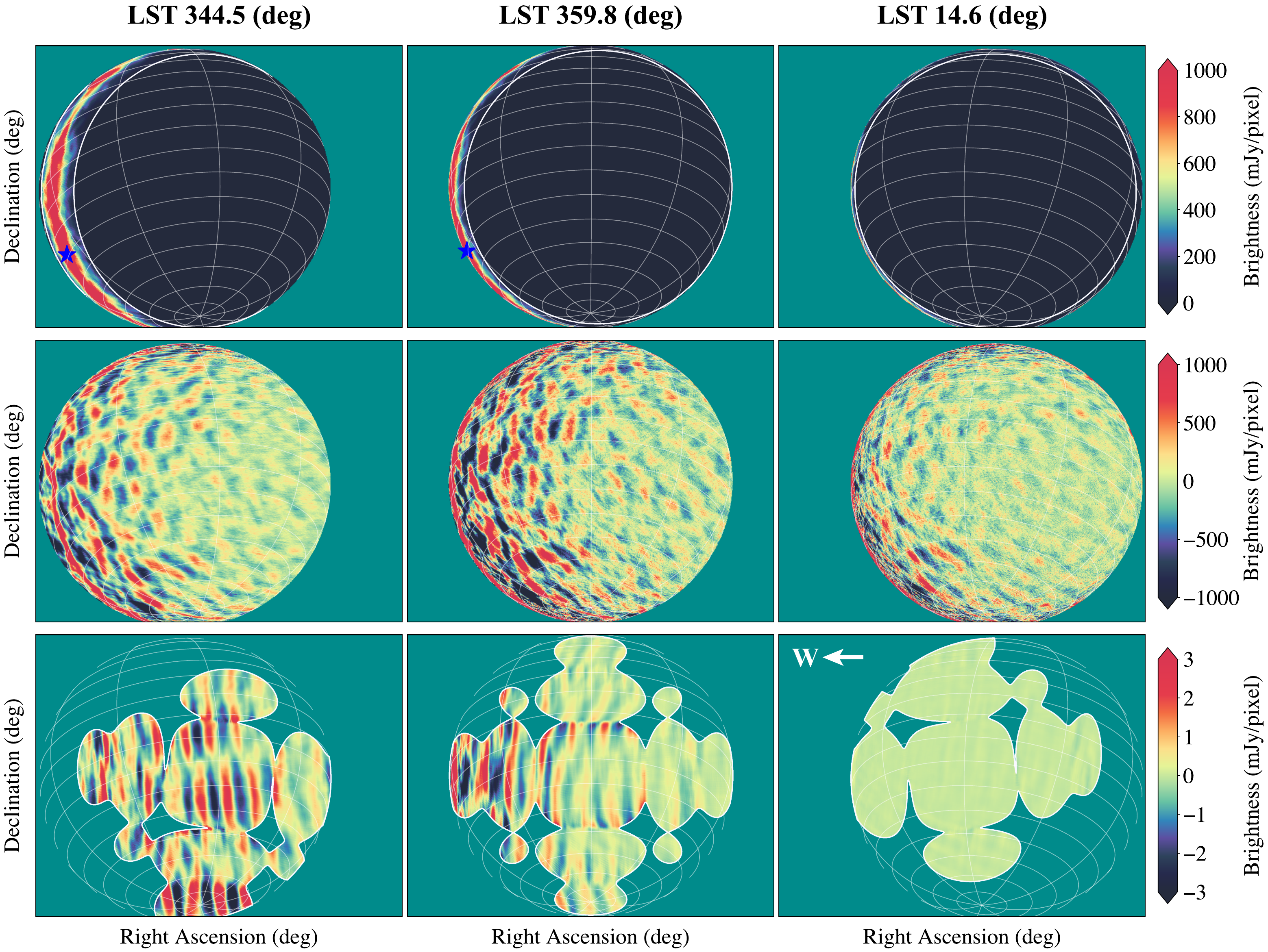}
	\caption{Simulations of the Galactic plane in uniform weighting. Each column corresponds to a different LST: 344.5\,deg (left), 359.8\,deg (middle), and 14.6\,deg (right). The top row is the input catalog from \citet{kriele_imaging_2022} out to $\pm$\,15\,deg in Galactic coordinates (white contours), with the Galactic center highlighted (blue star). The middle row is the input catalog as seen by an unpolarized interferometer with the MWA layout and uniform beam ($A(l,m,n) = 1$). The bottom row is the input catalog as seen by the MWA in N--S polarization out to 1\% beam attenuation (white contours). As the brightest part of the Galactic plane (top row, blue star) sets over the horizon, it aliases over the entirety of the image as seen by an interferometer (middle row). This is attenuated by the instrumental beam, but even a beam value of 1\% at the horizon is enough to contaminate most of the image (bottom row).}
	\label{fig:aliasing_panel}
\end{figure*}

We show the Galactic plane catalog in the top row of Figure~\ref{fig:aliasing_panel} for three different LSTs. These LSTs represent the observations for before, during, and after setting of the Galactic center (blue star). We purposely choose to simulate only $\pm$\,15\,deg away from Galactic plane itself (white contours) to capture the vast majority of its brightness while leaving as much as the sky unaltered by simulation as possible. 

Mathematically, an unpolarized, unattenuated interferometer beam can be simply described as having $A(l,m,n) = 1$ in Equation~\ref{eq:me}. While this is unphysical, it helps to show the base level of contamination prior to instrumental complications. The middle row of Figure~\ref{fig:aliasing_panel} shows the simulation of the Galactic plane for this simple case (specifically with a MWA layout). While the Galactic plane is obviously present at similar levels compared to the catalog, visible aliasing is also present throughout the entirety of the image. This additional, unwanted signal is caused by a finite sampling in Fourier space of an infinite gradient due to sudden opaqueness of the horizon, further emphasized by the extremely bright and extended Galactic plane.

The level of aliasing depends on the location of the Galactic plane and the amount of brightness transitioning across the horizon. The aliasing is brightest as the majority of the Galactic plane and center transitions across the horizon (Figure~\ref{fig:aliasing_panel}, middle). The Galactic plane aliasing is still significant prior to setting of the Galactic center (left), but less in comparison. After the setting of the Galactic center (right), the aliasing diminishes significantly but is still present due to the remaining brightness of the Galactic plane on the horizon.  As the Galactic plane sets, the associated aliasing moves, albeit slightly. There can be some decoherence from averaging the LSTs together, and it is in this way that some of the aliasing can be mitigated.

The spatial scales affected by the Galactic plane aliasing translates to a few megaparsecs to tens of megaparsecs at a redshift of 6.8. Estimates for EoR bubble sizes at that redshift are in the range of 10\,cMpc \citep{wyithe_characteristic_2004} up to 100\,cMpc \citep{lin_distribution_2016}, which could pose extraction issues for EoR tomography if no spectral or time mitigation is in place to disentangle Galactic plane aliasing from the EoR signal.

\subsection{Realistic interferometer}

Beam attenuation and polarization will affect the perceived level of Galactic plane aliasing. We show the MWA as an example, which has a large field of view (FoV). For an observation at an LST of 359.6\,deg (Figure~\ref{fig:aliasing_panel}, middle), the horizon that the Galactic plane sets over is attenuated to 0.2\% by the beam response.  

The Galactic plane is so bright, however, that this extreme attenuation does not remove the aliasing, apparent in the bottom row of Figure~\ref{fig:aliasing_panel}. The aliasing is present at the level of a few mJy~pixel$^{-1}$ at resolutions of 3$^\prime$.36~pixel$^{-1}$. As investigated in Section~\ref{sec:data}, this poses a problem in real data. 

The polarization of the instrument also changes the perceived shape of the aliasing given the location of the Galactic plane as it sets. In the example of Figure~\ref{fig:aliasing_panel}, the N--S aligned dipole beam was chosen, which is most sensitive E--W due to the toroidal electromagnetic response of a dipole. This polarization is most sensitive to the vertically aligned Galactic plane towards the E--W, creating vertical structures in the aliasing. The E--W aligned dipole beam, in contrast, will be most sensitive N--S, and thus will be less affected (see Figure~\ref{fig:data_panel} for an example).

\vspace{3em}

%%%%%%%%%%%%%%%%%%%%%%%%%%%%%%%%%
\section{Real Data Mitigation}
\label{sec:data}

%The Galactic plane and its aliasing can be seen in 2\,min of data with the MWA given a highly precise data analysis described in Section~\ref{subsec:framework}.

\begin{figure*}
\centering
	\includegraphics[width =\textwidth]{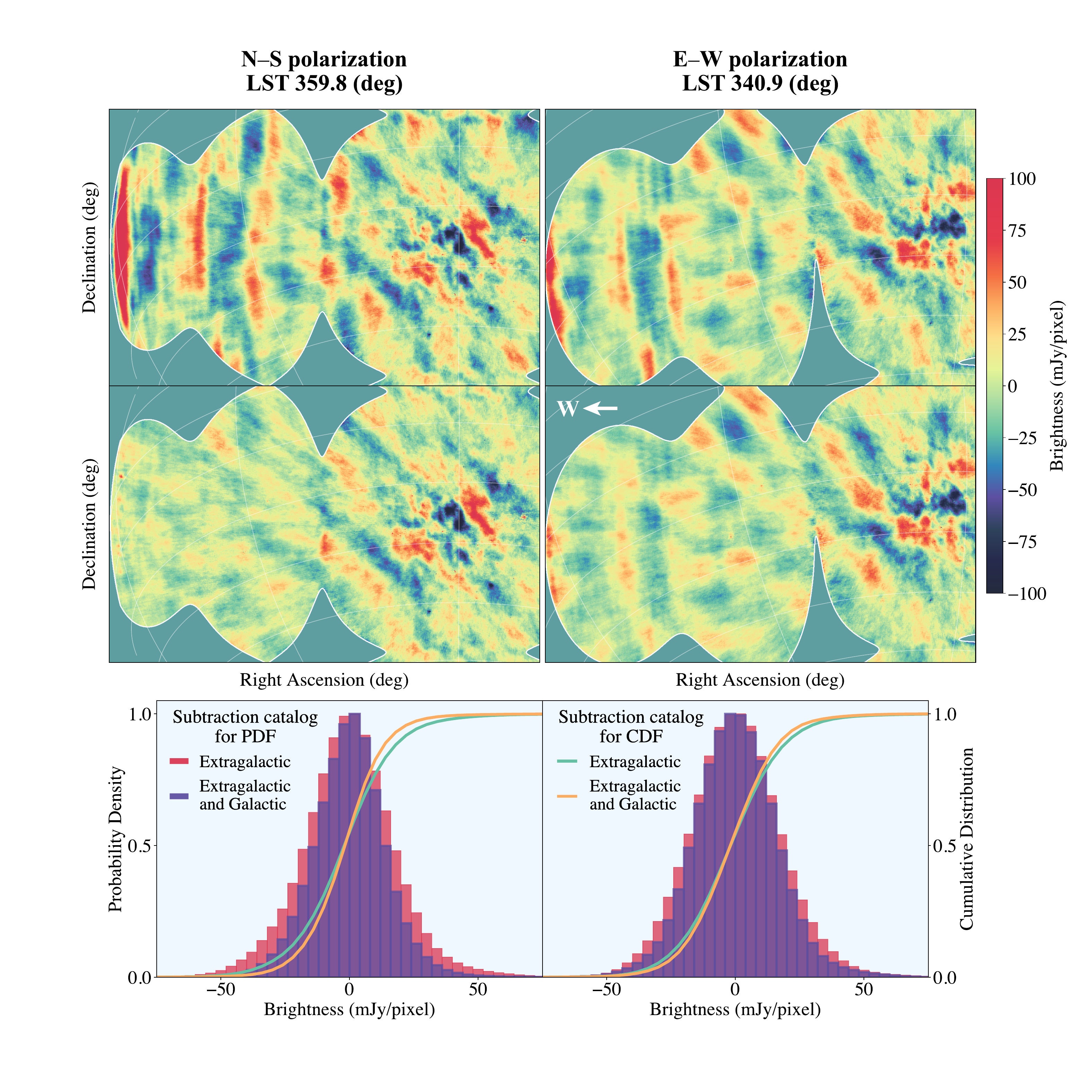}
	\caption{Removal of the Galactic plane in MWA data. Each column corresponds to a different LST and polarization: 359.8\,deg and N--S (left) and 340.9\,deg and E--W (right). The top row is a 2\,min observation where the extragalactic contribution has been removed, imaged with robust weighting out to the second sidelobe. Galactic plane aliasing is observed as vertical streaking. The middle row is the same observation with the Galactic contribution removed as well, reducing the observed aliasing. The bottom row is the probability distribution function (PDF) and cumulative distribution function (CDF) for the extragalactic-removed image (red histogram, cyan line) and the extragalactic-and-Galactic-removed image (purple histogram, yellow line). The statistics of the image where the Galactic contribution has been removed is more noise-like.}
	\label{fig:data_panel}
\end{figure*}

We now apply our knowledge of the Galactic plane to real data from the MWA. We start with a residual data image after attempting to remove extragalactic sources, shown in Figure~\ref{fig:data_panel} (top row) in Briggs 0 weighting. The remaining flux density on the sky is due to (1) mis-subtractions of extragalactic sources, (2) off-plane Galactic diffuse emission, and (3) the Galactic plane and its associated aliasing.

The Galactic plane is over-saturated in the N--S polarization for a zenith-pointed observation (left, top row), and just barely present in the E--W polarization for an observation two pointings before zenith (right, top row). The aliasing, observed as vertical streaking, is apparent in both polarizations and can easily be seen in the primary lobe for the N--S polarization. Much of the remaining flux density in the image is due to off-plane Galactic diffuse emission, which has not been subtracted in order to avoid contaminating the aliasing contribution.

Our simulations of the Galactic plane in Section~\ref{sec:sims} can also be used to subtract its contribution from the data to a significant degree. This additional subtraction is applied to the second row of Figure~\ref{fig:data_panel}. A drastic reduction in brightness can be seen not only at the location of the Galactic plane, but also throughout the entirety of the image due to a successful reconstruction of the aliasing. The aliasing structure is similar in size and shape to the remaining off-planar Galactic diffuse emission, suggesting that mapping the diffuse structure would be exceedingly difficult without aliasing removal. 

We show the change in noise statistics when the Galactic plane and its horizon aliasing are removed from the data in the bottom row of Figure~\ref{fig:data_panel}. The probability distribution function (PDF) and cumulative distribution function (CDF) of the residual brightness should follow Gaussian noise if all sources of sky brightness are removed from the image. The PDF and CDF of the extragalactic-subtracted image do appear Gaussian-like (red histogram and cyan line, respectively). However, when the Galactic plane is also subtracted from the image, the PDF and CDF are more Gaussian (purple histogram and yellow line, respectively) and have far less variance. This indicates that the Galactic plane and its associated horizon aliasing contributes excess variance to the image. 

\renewcommand{\arraystretch}{1}
\begin{table}

\centering \textbf{Table 1} \\ \footnotesize{Statistics of the Residual Images Depending on the Subtraction Catalog}

\setlength\tabcolsep{3.6pt}

\begin{tabular}{c  c  c  c  c} 
\hline \hline
\multicolumn{2}{c}{Subtraction Catalog} & Standard Deviation & Skewness & Kurtosis \\
 % & Deviation & \multirow{-2}{*}{Skewness} & \multirow{-2}{*}{Kurtosis} \\
\hline 

& E & 20.8 & 1.67 & 14.3 \\ 
\multirow{-2}{*}{N--S} & E \& GP & 14.7 & -0.01 & 5.84 \\ \hline
& E & 18.4 & 0.24 & 1.52 \\ 
\multirow{-2}{*}{E--W} & E \& GP & 16.4 & 0.05 & 1.45 \\ \hline

\end{tabular}

\raggedright
\footnotesize \textbf{Note.} E: Extragalactic; E \& GP: Extragalactic and Galactic plane. The first row and second row are the N--S image and E--W image in Figure~\ref{fig:data_panel}, respectively. Units are in mJy~pixel$^{-1}$.
%\end{TableNotes}

\label{table:stats}

\end{table}

Table~\hyperref[table:stats]{1} summarizes the statistics of the residual images. The standard deviation decreases by as much as 30\% in the N--S polarization when the Galactic plane and its aliasing are removed. Almost all skewness, or asymmetry, is gone. The resulting N--S image is also less tailed by almost 60\%, indicating that the distribution has fewer extremes and is closer to a Gaussian. The residual E--W image also experiences reductions in non-Gaussian behavior, albeit to a lesser degree. 

\vspace{3em}
%%%%%%%%%%%%%%%%%%%%%%%%%%%%%%%%%
\section{Consequences in Precision Science}
\label{sec:consequences}

We show how the setting of the Galactic plane affects two precision science cases: cataloging of faint, extragalactic radio sources and the 21\,cm EoR power spectrum. %To generate a base level of understanding, we do not include the effects of calibration.

\subsection{Extragalactic Catalogs}
\label{subsec:ec}

%The radio sky is filled with sources, including active galactic nuclei, relics and halos, high-redshift galaxies, and even some sources with unknown progenitors (i.e. ORCS, \citealt{norris_unexpected_2021}). 
%Population studies of radio sources require contribution from widefield telescopes, which are inherently affected by the setting of the Galactic plane. 

The process of making catalogs of extragalactic sources for population studies results in a well-calibrated, mosiac image with all known corrections applied. This image is then used to perform source finding --- algorithms which estimate a background rms to then pull out peaks which are over a supplied threshold, fitting Gaussians, wavelets, or points to build a model of the structure on the sky. Fainter sources are harder to capture in source-finding algorithms, and can be subject to Eddington bias, resolution bias, and sensitivity bias. 

However, the contribution of Galactic plane aliasing in a multi-observation mosiac over a range of LST introduces its own form of bias. Large-scale structure on the sky is artificially subsumed into the background RMS calculated via source-finding algorithms. This can bias both the total number and integrated flux of recovered sources.

These biases can be quantified via a completeness metric (e.g., \citealt{williams_lofar_2016,franzen_source_2019,hale_radio_2019}). False sources are injected into the final image which represent a theoretical population of complex sources that follow flux, shape, and positional distributions of known sources. The false sources are then recovered by the source-finding algorithms to quantify the level of flux density that is required to be statistically significant in the image.  

We closely follow the methods of \citet{lynch_mwa_2021} in calculating the completeness metric, with one exception. We inject the theoretical sources into our simulated images and compare the resulting completeness metrics between image simulations which include the Galactic plane and those that do not. The change in the completeness metric is representative of the effect that the Galactic plane aliasing has on faint-source detection. We do not include effects of calibration in order to investigate the base level of bias caused by the Galactic plane.

We generate 100 realizations of injected sources and investigate the bias in both number and flux density in their recovery in a simulated, 2\,hr image. When the Galactic plane and its aliasing is present in the image, 0.5\% more faint sources are recovered. This seemingly counter-intuitive result is due to the rms calculation of the source-finding algorithms subsuming the Galactic plane aliasing.

The Galactic plane aliasing is easily averaged away when large areas of image are used to calculate an underlying rms. However, when techniques like adaptive scaling are used to calculate a background rms, the size scale used can vary. In these cases, Galactic plane aliasing can artificially boost the calculated rms in smaller box sizes, which is subsequently subtracted from the image. This can reveal fainter sources in our idealized simulations. Over 95\% of sources which are revealed when the Galactic plane aliasing is subsumed into the rms calculation are less than 200\,mJy.

In general, our calculated bias is low enough to not cause much concern for extragalactic catalog science. This is in part due to the help of integration, which can mitigate the Galactic plane aliasing component. Currently, environmental and instrumental variations also cause far more bias (e.g., see \citealt{lynch_mwa_2021}).

\subsection{The 21\,cm EoR Power Spectrum}
\label{subsec:ps}

\begin{figure*}
\centering
	\includegraphics[width =\textwidth]{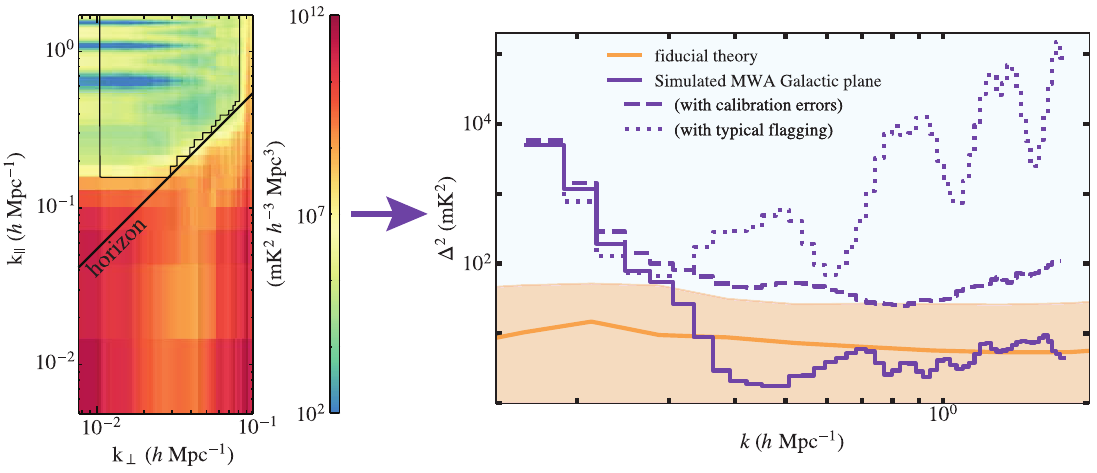}
	\caption{The reconstructed 2D (left) and 1D (right) power spectra (PS) of two hours of simulated Galactic plane data with the MWA. There are no other sources of power or noise. The contour on the 2D PS highlights bins used to generate 1D PS. The simulated Galactic plane power (solid purple) is just below the theoretical EoR signal (orange line with 95\% confidence region). Basic frequency-dependent calibration errors from an extragalactic calibration cause excess contamination (dashed purple). Not accounting for missing/flagged data is particularly harmful (solid purple), given that the Galactic plane along the horizon line is coupled to the missing/flagged modes.}
	\label{fig:ps}
\end{figure*}

The EoR power spectrum (PS) measurement relies on the natural separation of foregrounds from the 21\,cm signal in Fourier space. Noise-like signals which do not decorrelate with time could potentially preclude the measurement. 

PS analysis typically generates either a measured PS (e.g., delay or gridded delay, \citealt{parsons_per-baseline_2012}) or a reconstructed PS (e.g., \{$k_\perp$, $k_{||}$\}, \citealt{morales_toward_2004}). \citet{thyagarajan_confirmation_2015} first measured the implications of widefield foregrounds, including the Galactic plane, on measured PS analyses. In this work, we demonstrate the effects of the Galactic plane setting with simulations in a reconstructed PS analysis.

The left panel of Figure~\ref{fig:ps} shows the 2D PS of simulations of the Galactic plane through the MWA for a total of 2\,hr centered on field EoR0. The location of the foregrounds on the sky will affect the associated location of power in the PS space. As the foregrounds move away from the main FoV towards the horizon, they will move as a $k_\perp \propto k_{||}$ line, where the intercept increases with distance from phase center (e.g., \citealt{morales_four_2012, pober_importance_2016}). This line is shown in Figure~\ref{fig:ps} (left) as a solid black line for foregrounds at the horizon. Foregrounds are generally confined to the region below the solid black line, as is characteristic with the expected chromaticity of the instrument \citep{datta_bright_2010,morales_four_2012,parsons_per-baseline_2012,trott_impact_2012,vedantham_imaging_2012,hazelton_fundamental_2013,pober_opening_2013,thyagarajan_study_2013,liu_epoch_2014}. 

Our simulations model the data from the MWA. As such, we include regular flagging in frequency due to aliased channels from the polyphase filter bank, which are then averaged in frequency. The foregrounds are coupled to the channels with incomplete, missing data. This manifests as contaminated lines in $k_{||}$. The original locations of foregrounds are reflected in the contaminated $k_{||}$ modes, creating double-pronged features with the Galactic plane which can be seen in the left panel of Figure~\ref{fig:ps}. This is reminiscent of the ``pitchfork" effect (see \citealt{thyagarajan_confirmation_2015}).

We cylindrically average the 2D PS within the contoured region of the left panel of Figure~\ref{fig:ps} to investigate how the contamination propagates to the 1D PS. The orange line and shaded area in the right panel shows the expected EoR measurement with 95\% confidence \citep{barry_improving_2019}. The double-pronged features of the missing channels in the simulated Galactic plane (dotted purple) preclude the EoR signal in all modes in 1D space.

If there were no missing or incomplete channels, then the Galactic plane simulation is below the EoR signal (solid purple)\footnote{Remaining power above $k\approx0.4\,\textit{h}$\,Mpc$^{-1}$ is most likely due to unrelated analysis systematics in the beam model during the generation of the model visibilities.}. The effects of missing or incomplete channels can be mitigated by employing extensive removal techniques or full covariance weighting (e.g., \citealt{offringa_impact_2019}). 

We also show the consequences of excluding the Galactic plane in basic calibration in the 1D PS (dashed purple). We use the per-frequency gain solutions from a linear least-squares solver between (1) visibilities including extragalactic sources and the Galactic plane, and (2) visibilities from only extragalactic sources. This type of calibration is generally the starting point for most modern calibration techniques.

We exclude baselines less than 50\,$\lambda$ from the calibration \citep{patil_systematic_2016}, a common technique to avoid model dependence on diffuse emission. Nevertheless, the Galactic plane aliasing still affects the calibration such that there is at least an order of magnitude contamination in modes greater than 0.4\,\textit{h}\,Mpc$^{-1}$. This is due to a significant amount of power from this diffuse structure being greater than 50\,$\lambda$ \citep{byrne_map_2022}, and the extreme amount of precision required in this basic, per-frequency calibration formalism \citep{barry_calibration_2016}. Most calibration procedures are now much more advanced -- for example, we use an auto-correlation formalism in calibrating our data in Section~\ref{sec:data}.

Figure~\ref{fig:ps} focuses solely on simulations. However, we can also briefly investigate the effects on the EoR PS using real data. Using 2\,hr of data centered on EoR0 as an example, we generate limits on the 21\,cm signal. We use the same data selection and binning techniques as \citet{barry_improving_2019} for simplicity. The data was processed in two ways---with and without subtracting the Galactic plane. The calibration and baseline selection was kept the same to create a fair comparison. When the Galactic plane is removed in addition, there is a reduction in the upper limit by 0.4\% in the N--S polarization at 0.2\,\textit{h}\,Mpc$^{-1}$, which was the lowest $k$-mode in \citet{barry_improving_2019}. This reduction, whilst seemingly small, is still over 20x larger than the 21\,cm signal itself. Future work with real MWA data centered on EoR0 will investigate differences in data on a larger scale.

%%%%%%%%%%%%%%%%%%%%%%%%%%%%%%%%%
\section{Conclusion}
\label{sec:conclusion}

Widefield radio interferometry can be affected by aliasing inherent to the measurement itself from the bright, Galactic plane setting over the opaque horizon. We show simulations of this effect in great detail for the first time using the diffuse catalog from \citet{kriele_imaging_2022} as input into \texttt{WODEN}, a GPU-enabled analytic simulator \citep{line_woden_2022}, to create visibilities that are imaged with FHD \citep{sullivan_fast_2012, barry_fhd/eppsilon_2019}. These simulations, when subtracted from data, show improved image quality in just two minutes of observation. 

The aliasing of the Galactic plane setting is heavily attenuated by the beam. For the MWA, the aliasing structure can be upwards of tens of mJy per pixel in the primary lobe using Briggs 0 weighting. This structure moves slowly with LST, and thus will decorrelate slightly when integrated over a range of LST. 

The noise statistics of a real data image are improved with the removal of the Galactic plane and its aliasing. The resulting MWA image has less variance, skewness, and kurtosis. For future measurements of higher-order moments of the EoR \citep{watkinson_impact_2015, kittiwisit_measurements_2022}, like the skew spectrum \citep{ma_skew_2023,cook_impact_2024}, this could have a huge potential impact if done in image space.

We demonstrate the potential for impact on two current science goals: extragalactic cataloging and 21\,cm EoR power spectra detection. 
\begin{enumerate}

\item For extragalactic cataloging, the Galactic plane aliasing can be subsumed into the calculated rms maps of the mosaic image. This artificially boosts the subtracted rms, changing the amount of detected faint sources and their flux density. This is, in general, small enough to not be of concern when a range of LST are used to create the mosaic image. For example, with 2\,hr of MWA simulation, we see a difference of less than 0.5\% recovered sources, and changes in flux densities of recovered sources of less than 200\,mJy.
\item For 21\,cm EoR power spectra detection, there is quite a substantial effect near the horizon line in $\{k_\perp, k_{||}\}$ as expected. If there is regular flagging in frequency space, as is the case for the MWA, this shape will be reflected, creating mirrors of the horizon line that occupy a large number of $k$ modes. Basic per-frequency calibration formalisms will be affected by the Galactic plane aliasing, even if short baselines are excluded. Reconstructed PS analyses must consider the Galactic plane if it is setting, or at least employ a variety of techniques to remove its effects in flagging, calibration, and foreground power.
\end{enumerate}

Our simulations and real data analysis show that the setting of the Galactic plane can introduce aliasing into both Fourier-space and image-space statistical metrics. Given the brightness of the Galactic plane, beam attenuation of 0.1\% is not enough to remove this effect from some precision science goals.

\begin{acknowledgments}
The authors acknowledge the Wajarri Yamaji as the traditional custodians of the MWA site, Inyarrimanha Ilgari Bundara. N.B. acknowledges the support of the Forrest Research Foundation, under a postdoctoral research fellowship. The International Centre for Radio Astronomy Research (ICRAR) is a Joint Venture of Curtin University and The University of Western Australia, funded by the Western Australian State government. This research was supported by the Australian Research Council Centre of Excellence for All Sky Astrophysics in 3 Dimensions (ASTRO 3D), through project number CE170100013. This work was supported by resources awarded under Astronomy Australia Ltd’s merit allocation scheme on the OzSTAR national facility at Swinburne University of Technology. OzSTAR is funded by Swinburne University of Technology and the National Collaborative Research Infrastructure Strategy (NCRIS). This work was supported by resources provided by the Pawsey Supercomputing Research Centre with funding from the Australian Government and the Government of Western Australia.
\end{acknowledgments}


\vspace{5mm}
\facilities{MWA}


% \software{astropy \citep{2013A&A...558A..33A,2018AJ....156..123A},  
%           Cloudy \citep{2013RMxAA..49..137F}, 
%           Source Extractor \citep{1996A&AS..117..393B}
%           }


%\appendix
%%%%%%%%%%%%%%%%%%%%%%%%%%%%%%%%%



\bibliography{example}{}
\bibliographystyle{aasjournal}



\end{document}